\documentclass{aa}

\usepackage{graphicx}
\usepackage{txfonts}
\usepackage{hyperref}
\hypersetup{
    colorlinks=true,
    linkcolor=blue,
    citecolor=blue
    }
\begin{document}

   \title{Supermassive black hole formation via collisions in black hole clusters}


    \author{B. Gaete  \inst{1} \and D.R.G. Schleicher\inst{1} \and  A. Lupi\inst{2,3} \and B. Reinoso\inst{4} \and M. Fellhauer\inst{1} \and M. C. Vergara\inst{5}  }

    \institute{
        $^1$Departamento de Astronom\'ia, Facultad Ciencias F\'isicas y Matem\'aticas, Universidad de Concepci\'on, Av. Esteban Iturra s/n, Barrio Universitario, Concepci\'on, Chile \email{begaete@udec.cl}  \\ 
        $^2$Dipartimento di Fisica G. Occhialini, Università di Milano-Bicocca, Piazza della Scienza 3, I-20126 Milano, Italy \\
        $^3$INFN, Sezione di Milano-Bicocca, Piazza della Scienza 3, I-20126 Milano, Italy\\
        $^4$Universit\"at Heidelberg, Zentrum f\"ur Astronomie, Institut f\"ur theoretische Astrophysik, Albert-Ueberle Str. 2, 69120 Heidelberg, Germany \\
        $^5$Astronomisches Rechen-Institut, Zentrum für Astronomie, University of Heidelberg, Mönchhofstrasse 12-14, 69120, Heidelberg, Germany
             }

   \date{Received September 15, 1996; accepted March 16, 1997}

 
  \abstract
 {More than 300 supermassive black holes have been detected at redshifts larger than $6$, and they are abundant in the centers of local galaxies.    {Their formation mechanisms, however, are still rather unconstrained.}  A possible origin of these supermassive black holes could be through mergers in dense black hole clusters, forming as a result of mass segregation within nuclear star clusters in the center of galaxies. In this study, we present the first systematic investigation of the evolution of such black hole clusters where the effect of an external potential is taken into account. Such a potential could be the result of gas inflows into the central region, for example as a result of galaxy mergers. We show here that the efficiency for the formation of a massive central object is mostly regulated by the ratio of cluster velocity dispersion divided by the speed of light, potentially reaching efficiencies of $0.05-0.08$ in realistic systems. Our results show that this scenario is potentially feasible and may provide seeds black hole  of at least $10^3$~M$_\odot$. We conclude that the formation of seed black holes via this channel should be taken into account in statistical assessments of the black hole population.}


   \maketitle
%
\section{Introduction} \label{section:Introduction}

The existence of supermassive black holes (SMBHs) and their physical nature has been confirmed through different independent observations, including the orbits of the S2 stars near the center of Milky Way with the GRAVITY instrument \citep{gravity2018}, as well as the observation of their shadows at the centers of M87 and Sagitarius A* \citep{eht2019,eht2022}. Observed through the detection of Active Galactic Nuclei (AGN) at high redshift \citep[e.g.][]{Shankar2010}, even at redshifts larger than  $ z > 6$, more than 300 quasars have been detected \citep[e.g.][]{Banados_2016,Inayoshi2020,XFan+2023}. These objects are very rare with number densities of $\thicksim 1 \; \mathrm{Gpc}^{-3}$ and have been found so far in optical/infrared(IR) surveys that cover a large portion of the sky, such as The Sloan Digital Sky Survey (SDSS), the first survey to discover a high-redshift quasar \citep{fan2001}. They are common in the centers of local galaxies \citep[e.g.][]{Ferrarese2000, Tremaine2002, Gueltekin2009} and their masses are in the range of $10^{6}-10^{10} M_{\odot}$.

The most distant quasar detected so far was discovered by the James Webb Space Telescope (JWST) at redshift $z \approx 10.3$ magnified by the cluster Abel 2744, the SMBH has a mass of $ \approx 10^{7} - 10^{8} \mathrm{M}_{\odot}$ assuming accretion at the Eddington limit \citep{Goulding_2024}.

In the local Universe the rarest SMBHs are the so-called ultra-massive ones; over the last decade observations have established the existence of a few of these with masses $\gtrsim 10^{10} \mathrm{M}_{\odot}$ in some bright cluster galaxies \citep[e.g.][]{McConnell2011,Hlavacek2012,Wu2015,Schindler_2020}. 


In the local Universe, galaxies were also found to host nuclear star cluster (NSCs) at their centers \citep{Neumayer2020}. The most massive NSCs are the densest known stellar systems and can reach mass surface densities of $\approx 10^{6} 
\: \mathrm{M}_{\odot}/\mathrm{pc}^{2}$ or higher. Some important features of these objects and an important topic to study are their correlations with properties of their host galaxies, such as  the tight correlations with the masses of their host galaxy \citep{Wehner2006, Rossa2006, Ferrarese2006}. The correlation between the mass of the spheroidal component of the host galaxy and the mass of the SMBH, as well as that with the bulge velocity dispersion, is another crucial aspect to consider \citep{G_ltekin_2009,Magorrian_1998}.There are a number of cases where NSCs and SMBHs were found to co-exist \citep{Filippenko2003, Seth2008, Graham2009, Neumayer2012, Nguyen2019}. Other nearby examples of SMBH detections within NSCs are M31 \citep{Bender2005}, M32 \citep{Verolme2002, Nguyen2018}, NGC 3115 and the Milky Way \citep{Tonry1984, Dressler1988, Richstone1990, Kormendy1992, Marel1994}. The co-existence suggests that the build-up of NSCs and the growth of SMBHs are closely related \citep[see also][]{Escala2021, Vergara2022}. The high masses of the SMBHs at an early age of the Universe where we observe these objects are a real challenge for theories of their formation. If we assume a constant accretion at the Eddington limit with only $10\%$ of the matter falling into the black hole (BH) being radiated away, a stellar-mass black hole with a mass of $ = 10 \; \mathrm{M_{\odot}}$ requires a timescale of  $\mathrm{t}_{accr} \approx 1 \; \mathrm{Gyr}$ to reach the masses of SMBHs observed in the most massive AGN \citep{Shapiro2005}. However, it is unlikely to grow so much  because of the removal of the gas reservoir by UV radiation and supernova (SN) explosions of the Pop III stars in the shallow gravitational potential wells of minihalos \citep{Johnson_2007,Whalen_Oshea_2008,Milosavljevic_2009}. This suggests that the seed black hole must have formed at redshift $ z \geq 15 $ with a mass of $\approx 10^{5}$~M$_\odot$, or the black hole seed had a lower mass but accreted very efficiently, or a combination of both.

One promising formation scenario is the Direct Collapse (DC) \citep{latif2013, latif2015}, which involves the gravitational collapse of massive gas clouds in atomic cooling haloes ($T_{vir} > 10^{4}$ K, M $\approx {10^{8}}$ M$_{\odot}$) at high redshift \citep{wise2019}. This process was suggested to form SMBH seeds with masses around ${10^{3-5}}$~M$_{\odot}$. However, the presence of molecular hydrogen could lead to cloud fragmentation \citep{omukai2008, suazo2019}, preventing the formation of massive objects. Another scenario is associated with the dynamics within stellar clusters. Fragmentation at high density may give rise to the formation of ultra dense clusters \citep{omukai2008, Devecchi_2009}. Due to its high stellar density, this cluster can undergo runaway core collapse in a short time, forming a central intermediate-mass black hole (IMBH) with a mass of approximately $10^{2-4}$~M$_{\odot}$ \citep[e.g.][]{Zwart2004,Sakurai2017,Reinoso2018, Reinoso2020, Vergara2021, Vergara2023}. In this scenario, a newly born dense star cluster could still be embedded in gas, aiding in the formation of a massive black hole  seed through the inflow of gas into the cluster \citep{Tagawa_2020}. This process increases the gravitational potential of the cluster, reduces the escapers{, which we define as BHs with more kinetic than potential energy}, and deepens the potential well of the cluster. Furthermore, in this scenario, the protostar may accrete gas, increasing its radius and, consequently, its cross section. Gas dynamical friction can drive a more efficient core collapse \citep[e.g.][]{Tagawa_2020, Schleicher2022}.

Dynamical friction can also cause massive objects to sink to the center of a star cluster, where at the end of their lifetime the massive objects will evolve into stellar mass black holes or neutron stars, forming a dark core. The dynamical evolution of such black hole clusters has been examined by \citet{Quinlan1987, Quinlan1989}, finding that for typical parameters these clusters will dissolve due to the ejection of black holes as a result of three-body interactions \citep[see also][]{Chassonnery2021}. A solution to this problem has been proposed by \citet{Davies2011}, showing that gas inflows after galaxy mergers could steepen the potential of the cluster, increase the velocity dispersion, and reduce the timescale for contraction due to gravitational wave emission in comparison to the three-body ejection timescale. \citet{Lupi2014} investigated the statistical implications of such a scenario by implementing it into a semi-analytic model of galaxy evolution, showing that this formation channel may contribute a substantial amount of seed black holes. In an independent investigation, \citet{Kroupa2020} found that the steepening of black hole clusters through inflows of gas could explain the presence of supermassive black holes in high-redshift quasars. 

In this paper, we investigate this scenario in more detail, exploring the evolution of a black hole cluster in an external potential to determine under which conditions it can lead to the formation of an IMBH, as well as the efficiency of that process. In Section (\ref{section:Model}), we describe the model that forms the basis of our simulations. Section (\ref{section:Simulations}) outlines the methodology employed, along with the initial conditions for the simulations. In Section(~\ref{section:Results}), we present the results of the evolution of the clusters, detailing the influence of the external potential and the post-Newtonian effects on the formation of massive objects. We explore mergers via gravitational waves, the influence of escapers, and various properties of the binaries formed within the cluster. In Section (\ref{section:Discussions}) we provide a discussion of neglected effects including possible considerations for future research. Finally, in Section (\ref{section:Conclusions}), we present our conclusions.

\section{Model} \label{section:Model}

The theoretical framework of this project is a variation of the model introduced in the previous section on runaway mergers in dense star clusters. The model considers mergers in dense black hole clusters, following the framework of \citet{Davies2011}. Due to mass segregation, the stellar mass BHs are assumed to have sunken to the center of the core of a nuclear star cluster. In stellar systems there is a  tendency towards equipartition of kinetic energies, so the most massive objects will tend to move more slowly on average and then massive objects fall deep into the potential well, while light objects tend to move fast and move out, and may reach the velocity necessary to escape. This instability is known as the equipartition instability or Spitzer instability causing mass segregation, leading to the formation of a dark core.

We assume that the stars and other remnants in the core of the  cluster can be ignored as their individual masses are much smaller than those of the stellar mass BHs and thus they will be absorbed by the BHs or they may be pushed outside of the radius of the dark core \citep{Banerjee2011,Breen2013}. The cluster which is more than $50 \: \mathrm{Myr}$ old is assumed to consist of N equal mass stellar BHs each with mass $m_{BH}$. Some BH - BH interactions can lead to escapers but a significant fraction of the initial stellar mass BHs remains in the cluster \citep{Mackey2007}. { For simplicity we assume here that the BHs formed in the cluster do not grow during the initial 50 Myrs because of stellar feedback (radiation, winds, and SNe).}

Binaries within the dark core stabilize the cluster against core collapse as the binaries are a heating source \citep{Hills1975, Heggie1975, Miller2002}. Thus the dark core evolves as the BH population self-depletes through the dynamical formation of BH binaries in triple encounters which, after their formation, may exchange energy with a third BH. Some of these interactions could lead to BH escapers, though due the deep potential well, the cluster retains most of its BHs. According to the Hénon principle \citep{Henon1961, Henon1975}, the energy generation rate in the cluster core from encounters between single BHs/binaries  with hard binaries is regulated by the mass of the system. Such encounters transform binding energy into kinetic energy, which supports the cluster against core collapse. While soft binaries will be split by interactions in binary-single encounters,  hard binaries tend to harden in binary-single encounters. We introduce here the critical value of the semi-major axis describing the transition between soft and hard binary systems,

\begin{equation}\label{eq:soft_hard}
    a_{h/s} = \frac{Gm_1m_2}{<m>\sigma^{2}},
\end{equation}

where $m_1$ and $m_2$ are the masses of the primary and secondary of the binary system, $\langle m\rangle$ describes the average black hole mass in the cluster core and $\sigma$ the velocity dispersion. Binaries with a semi-major axes $a > a_{h/s}$ are then referred to as  soft binaries and will be disrupted due to gravitationally encounters, while only hard binaries with $a < a_{h/s}$ can survive. 
The timescale of a binary within a cluster to gravitational interact with another object is given by \citep{Binney2008}

\begin{equation} \label{eq:encounter_2+1}
    \tau_{2+1} \simeq 6 \times 10^{8} x \frac{M_{c,6}^{2}}{v_{\infty,10}^3}  \:\:\: \mathrm{yr},
\end{equation}

where $M_{c,6}$ is the total mass of the cluster in units of $10^{6} \: M_{\odot}$, $x$ is the ratio of binary binding  energy to  kinetic energy, and $v_{\infty,10}$ is the relative velocity at infinity in units of 10 $km/s$. In virial equilibrium we have $v_{\infty,10}$ $\approx$ 4.36 $\sqrt{ GM_{c}/r_{h}}$ \citep{Binney2008}, with $r_{h}$ is the cluster half-mass radius. Once the dark core reaches enough velocity dispersion (corresponding to very large density), the dynamical binaries formed in the cluster will be sufficiently tight will  quickly merge via gravitational wave (GW) emission, as then the time scale of GW emission will be equal to or shorter than the time scale of binary-single encounters. As the binding energy stored in the binaries is lost via GW emission, the binaries cease to be a source of heating for the cluster and  core collapse takes place. The decay time of a BH binary with an initial separation $a$ and eccentricity $e$ is \citep{Peters1964},

\begin{equation}  \label{eq:GW}
    \tau_{gw} \simeq 5 \times 10^{-3} c^{5} G \frac{m_{bh}}{v_{\infty,10}^{8}} x^{-4} (1-e^2)^{7/2} \:\:\: \mathrm{yr}.
\end{equation}

The gravitational binary-single interactions  will leave the binaries with a thermal distribution of the orbital eccentricities,  where the median eccentricity is $e_{med} = 1/\sqrt{2}$. This effect reduces the typical binary merger time by a factor $\approx 10$. If $\tau_{gr} < \tau_{2+1}$ binaries will merge avoiding the transfer of their binding energy to kinetic energy via gravitational interactions, lose the energy that is stored in the binaries and thus the binaries will not keep heating the cluster as a result. Then  the energy equilibrium breaks and core collapse is expected to happen. 

This scenario thus requires a mechanism to shrink the radius of the cluster and/or increase its mass. The dark core thus needs to become more dense, so that the black holes may merge via runaway processes and stay within the cluster. In the scenario proposed by \citet{Mayer2010}, the self-gravitating gas is subject to instabilities that funnel much of the low angular momentum gas to the center  to scales of $0.2 \; \mathrm{pc}$ or less. It is thus very efficient in contracting the core of the cluster, to increase the central densities and enhance the mass segregation, leading to fast interactions between stellar mass black holes that could lead to a quick coalescence and the formation of a massive BH seed. High resolution cosmological simulations of galaxy formation by \citet{Bellovary2011} show a gaseous inflow due to a combination of accretion of matter from the cosmic web-filaments and mergers of galaxies, providing a significant inflow of gas comparable to or greater than the stellar mass in the cluster at high redshift ($z > 10$). 

Independent of the primordial mass segregation the inflow of gas into the cluster will make the black hole cluster shrink given the steepening of the potential. This increases the interactions between the BHs, while the initial fraction of hard binaries also affects the re-expansion of the cluster due to their heating effect. In this scenario the gas only contributes to deepening the potential well, while we neglect here the dynamical friction that could make the cluster even more dissipative and further enhance the probability to form a very massive object, as well as the formation of the gas itself (cooling, fragmentation, star formation) \citet{Kroupa2020} have further investigated this scenario, defining the gas mass that falls into the black hole cluster $M_g = \eta_{g} N m_{BH}$. They find this scenario to be feasible for $0.1 \: < \eta_{g} \: < 1.0$  with $R_{vir } \lesssim 1.5 - 4 \:  \mathrm{pc}$ and a total BH mass in the cluster $M_{BH} \gtrsim  10^{4} \: M_{\odot}$, where the cluster could reach a relativistic state (1$\%$ speed of light) within much less than a $\mathrm{Gyr}$ , while for $\eta_{g} < 0.06 $ the BH cluster expands because the binary heating dominates over the gas drag. For large values such as $\eta_{g} > 6$ the black hole cluster may even be in the relativistic regime from the beginning.

\section{Simulations}  \label{section:Simulations}

To resolve the gravitational dynamics in the cluster, including post-Newtonian corrections, we use the \textsc{Nbody6++GPU} code \citep{Wang_2015}. \textsc{Nbody6++GPU} uses a Hermite $4^{th}$ order integrator method \citep{Makino_1991}. It also includes a set of routines to speed up the calculations such as using spatial and individual time steps and a spatial hierarchy which considers a list of neighbor particles inside a given radius to distinguish between the regular force and the irregular force \citep{Ahmad_1973}. In this version the gravitational forces are computed by Graphics Processing Units (GPUs) \citep{Wang_2015,Nitadori_2012}. It further uses an algorithm to regulate close encounters \citep{Kustaanheimo_1965}. Finally, the code includes post-Newtonian effects as described below \citep{Kupi_2006}.

As we saw above, \textsc{Nbody6++GPU} includes KS regularization \citep{Kustaanheimo_1965}, and this algorithm starts to operate when 2 particles are tightly bound, replacing them with one particle and treating their orbit internally. This scheme is modified to allow for relativistic corrections to the Newtonian forces by expanding the accelerations in a series of powers of $1/c$ \citep{Soffel1989}:

\begin{equation}\label{eq:PN_corrections}
    \underline{a} = \underbrace{\underline{a}_{0} }_{Newt.}  + \underbrace{c^{-2}\underline{a}_2}_{1\mathcal{PN}} +  \underbrace{c^{-4} \underline{a}_4}_{2\mathcal{PN}} + \underbrace{c^{-5} \underline{a}_5}_{2.5 \mathcal{PN}} + \underbrace{c^{-6} \underline{a}_6}_{3 \mathcal{PN}} + \underbrace{c^{-7} \underline{a}_7}_{3.5 \mathcal{PN}}   + \mathcal{O}(c^{-8}),
\end{equation}
where $\underline{a}$ is the acceleration of particle 1, $\underline{a}_0 = -Gm_2n/r^{2}$ is its Newtonian acceleration, and 1 $\mathcal{PN}$ , 2 $\mathcal{PN}$ and 2.5 $\mathcal{PN}$ are the Post-Newtonian corrections to the Newtonian acceleration,  where $1\mathcal{PN}$ and 2 $\mathcal{PN}$ correspond to the pericenter shift , 2.5 $\mathcal{PN}$ , 3 $\mathcal{PN}$ and  3.5 $\mathcal{PN}$ to the quadrupole gravitational radiation. The corrections are integrated into the KS regularization scheme as perturbations, similarly to what is done to account for passing stars influencing the KS pair \citep{Brem_2013}.

Finally the criterion for particle mergers is calculated from their Schwarzschild radii as

\begin{equation} \label{eq:merger_condition}
    |R_{i,j}|  < 10\frac{G}{c^{2}} (m_i+m_j),
\end{equation}
where $G$ is the gravitational constant, $c$ is the speed of light, $m_i$ and $m_j$ the mass of particles $i$ and $j$, with $|R_{i,j}|$ the distance between the particles in the binary system. This equation shows that the two BHs can merge only when their separation is smaller than 5 times the sum of their Schwarzchild radii. The mass of the new BH that forms is then  given by the sum of the masses of the two merging black holes, considering a ideal case where we neglect the mass loss due to GW radiation.

In this project, we use the model introduced in Sec. (\ref{section:Model}) to explore the evolution and the formation of an SMBH seed in the dark core of a NSC. We perform a range of simulations to study how the presence of an external gas potential affects a dark core, the contraction of the dark core and the growth of a SMBH seed via runaway mergers. The configurations that we consider to model the dark core of a NSC is a spherical cluster of $N = 10^{4}$  stellar mass black holes with identical BH masses of $m_{BH} = 10 \: M_\odot$ at the beginning of the simulations. The spatial distribution is an isotropic Plummer sphere \citep{plummer1911} in virial equilibrium with virial radius of $R_v = 1.7 R_{a}$ with  $R_{a} = 0.59$. The analytic potential is given by a Plummer distribution with a mass $M_{gas} = \eta_{g} M_{BH}$ and the same virial radius of the cluster,  where we vary the gas mass fraction of the cluster as $\eta_{g} = 0.0,0.1,0.3,0.5,1.0$. 

\begin{figure}[htbp]
    \centering
    \includegraphics[width=0.48\textwidth]{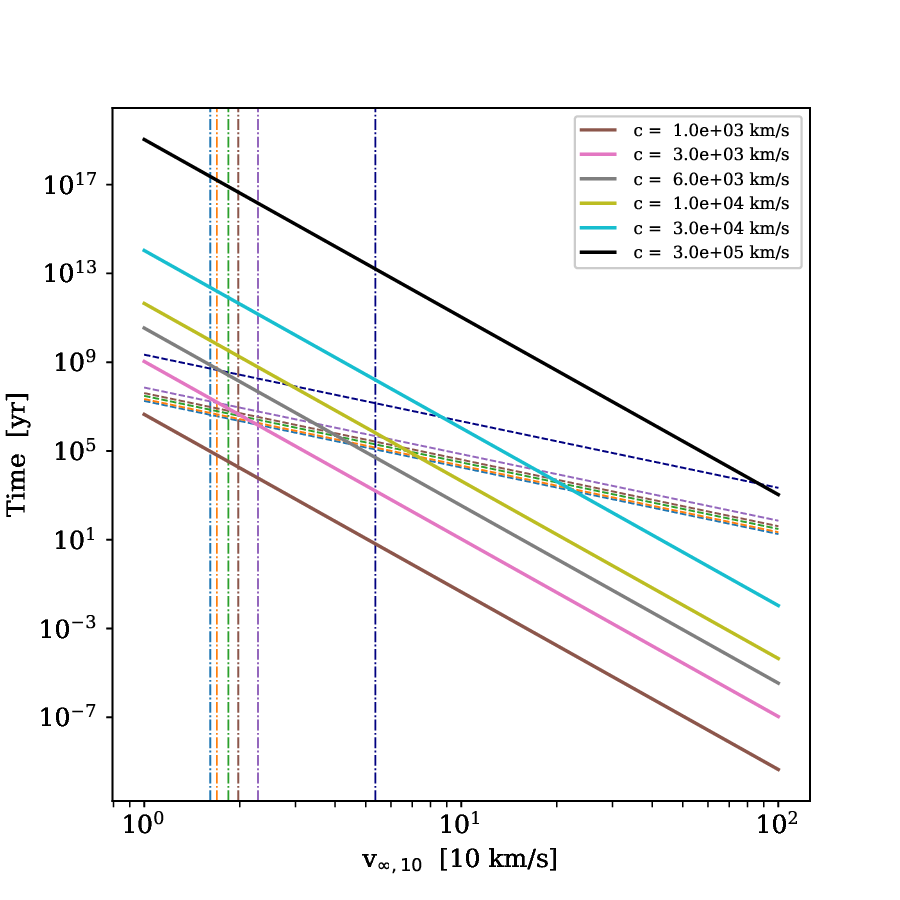}
    \caption{We show different time scales in years. The time scale for binary-single encounters given by Eq.~\ref{eq:encounter_2+1} (dashed lines)  and gravitational radiation inspiral given by Eq.~\ref{eq:GW} (solid lines), considering a variation of the gas mass fraction between $\eta_{g} = 0.0, 0.1, 0.3, 0.5, 1.0, 10.0$, with the lowest value in the blue dashed line and the highest  value in the violet. We also employ different values of the speed of light, from $c = 10^{3} \: \mathrm{km/s}$ (brown dashed line) to the real value of $c = 3 \times 10^{5} \: \mathrm{km/s}$ (dark solid line). The vertical lines show the velocity dispersion given by $v_{\infty}$  for different  gas mass fractions.}    
    \label{fig:1:t_vs_v}
\end{figure}

Modeling such a cluster employing the physical velocity of light $c = 3 \times 10^{5} \: \mathrm{km/s}$ is computationally unfeasible as too many iterations would be required until the binaries evolve into a state where the relativistic effects become important enough for the BH mergers to occur. Mergers via gravitational radiation are strongly dependent on the speed of light, as seen in Eq. (\ref{eq:GW}), and the time scale for gravitational wave emission is proportional to $c^5$. Besides increasing the time for mergers, it also increases the time to solve the equation of motion, because as we see above in the Hermite scheme we need to compute not only the acceleration but also the derivative, and we need to do this for every factor of the post-Newtonian corrections 1 $\mathcal{PN}$,2 $\mathcal{PN}$, 2.5 $\mathcal{PN}$, 3 $\mathcal{PN}$ and 3.5 $\mathcal{PN}$. Simulations considering the real speed of light could take years to model the systems considered here. Although a simulation employing the real speed of light may not be feasible. However, one can employ a reduce speed of light, which makes the calculations computationally affordable and allows to explore the behavior of the system. By exploring the dependence of the speed of light, we can then extrapolate the simulations outcome to the real value of c. We vary the speed of light as  $c = 10^{3}; 3 \times 10^{3}; 6 \times 10^{3}; 10^{4}, 3 \times 10^{4} \:\: \mathrm{km/s}$, which also affects the radii of the BHs in the cluster via the Schwarzschild radii and the criterion for the mergers. As we will see in our results, for values of the speed of light sufficiently high the dependence on this parameter in fact becomes relatively weak. The  time evolution of all clusters is considered  over a time of  $T =  1.4 \: \mathrm{Gyr}$. All configurations are given in Table~(\ref{table:1}). We conducted 4 simulations with different random seeds to ensure diverse initial conditions at the start of each simulation, thus corresponding to a total of 100 simulations. For these initial condition we estimate the range of root mean square (rms) velocities necessary for binary systems to efficiently merge via gravitational radiation before being ejected as a result of 2+1 encounters in  Fig. (\ref{fig:1:t_vs_v}). For the range of rms velocities in our simulations, the figure indeed shows that the timescale for gravitational wave emission becomes less than the timescale for 2+1 encounters for a speed of light of $c=3000$~km/s, thus potentially enhancing the formation of a very massive object in this regime, while for larger values of the speed of light the mergers will be delayed, and escapers due to 2+1 interactions could play a certain role. 


\begin{table}[htbp] 
\caption{Initial conditions of the simulations presented here. The initial amount of black holes in the cluster is $N$, the total BH mass in the cluster is $M_{BH}$, the fraction of gas mass in  the cluster is given by $\eta_{g}$, the virial radius is denoted as $R_v$ and the speed of light that we use in the simulation is given by $c$.}
\centering
\begin{tabular}{cccccc} 
\hline \hline  \\ 
IDs & N        & M$_{BH}$ {[}M$_{\odot}${]} & R$_v$ {[}pc{]} &  $\eta_{g}$ &  c {[}km/s{]} \\ \\
\hline 
\\
1   & 10$^{4}$ & 10$^{5}$               & 1.0            & 0.0               & 10$^{3}$         \\
2   & 10$^{4}$ & 10$^{5}$               & 1.0            & 0.1               & 10$^{3}$         \\
3   & 10$^{4}$ & 10$^{5}$               & 1.0            & 0.3               & 10$^{3}$         \\
4   & 10$^{4}$ & 10$^{5}$               & 1.0            & 0.5               & 10$^{3}$         \\
5   & 10$^{4}$ & 10$^{5}$               & 1.0            & 1.0               & 10$^{3}$         \\
6   & 10$^{4}$ & 10$^{5}$               & 1.0            & 0.0               & 3$\times$10$^{3}$ \\
7   & 10$^{4}$ & 10$^{5}$               & 1.0            & 0.1               & 3$\times$10$^{3}$ \\
8   & 10$^{4}$ & 10$^{5}$               & 1.0            & 0.3               & 3$\times$10$^{3}$ \\
9   & 10$^{4}$ & 10$^{5}$               & 1.0            & 0.5               & 3$\times$10$^{3}$ \\
10  & 10$^{4}$ & 10$^{5}$               & 1.0            & 1.0               & 3$\times$10$^{3}$ \\
11  & 10$^{4}$ & 10$^{5}$               & 1.0            & 0.0               & 6$\times$10$^{3}$ \\
12  & 10$^{4}$ & 10$^{5}$               & 1.0            & 0.1               & 6$\times$10$^{3}$ \\
13  & 10$^{4}$ & 10$^{5}$               & 1.0            & 0.3               & 6$\times$10$^{3}$ \\
14  & 10$^{4}$ & 10$^{5}$               & 1.0            & 0.5               & 6$\times$10$^{3}$ \\
15  & 10$^{4}$ & 10$^{5}$               & 1.0            & 1.0               & 6$\times$10$^{3}$ \\
16  & 10$^{4}$ & 10$^{5}$               & 1.0            & 0.0               & 10$^{4}$         \\
17  & 10$^{4}$ & 10$^{5}$               & 1.0            & 0.1               & 10$^{4}$         \\
18  & 10$^{4}$ & 10$^{5}$               & 1.0            & 0.3               & 10$^{4}$         \\
19  & 10$^{4}$ & 10$^{5}$               & 1.0            & 0.5               & 10$^{4}$         \\
20  & 10$^{4}$ & 10$^{5}$               & 1.0            & 1.0               & 10$^{4}$         \\
21  & 10$^{4}$ & 10$^{5}$               & 1.0            & 0.0               & 3$\times$10$^{4}$ \\
22  & 10$^{4}$ & 10$^{5}$               & 1.0            & 0.1               & 3$\times$10$^{4}$ \\
23  & 10$^{4}$ & 10$^{5}$               & 1.0            & 0.3               & 3$\times$10$^{4}$ \\
24  & 10$^{4}$ & 10$^{5}$               & 1.0            & 0.5               & 3$\times$10$^{4}$ \\
25  & 10$^{4}$ & 10$^{5}$               & 1.0            & 1.0               & 3$\times$10$^{4}$ \\
\hline \\
\end{tabular}
\label{table:1}
\end{table}

\section{Results}\label{section:Results}

In this section, we present the results of the simulations in which we explore the behavior of the black hole clusters, taking into account the influence of an external gas potential as well as variations in the ratio of gas mass over the total mass in BHs. Additionally, we consider the effects of varying the speed of light  and how it affects the evolution and growth of the central object. The setups we consider are detailed in Table (\ref{table:1}). In the next subsection, we will focus on three specific clusters with IDs 1, 5, and 25 as indicated in Table (\ref{table:1}), corresponding to clusters with low speed of light and no external potential, low speed of light and high external potential and a high speed of light with a high external potential.

\subsection{Evolution of black hole clusters}

The evolution  of a BH cluster from birth to the moment of core collapse is described by \citet{Spitzer1987}, where the cluster evolves toward the collapse via two-body relaxations at the half mass radius. This time tends to increase when the cluster is affected by a background potential \citep{Reinoso2020}  and is given by

\begin{equation}\label{eq:relaxation_time}
    t_{rh,ext} =  0.138\frac{N (1+\eta_{g})^{4}}{ln(\gamma N )} t_{cross,ext},
\end{equation}

where $N$ is the number of particles in the cluster, $\eta_{g}$ is the fraction of gas mass, $\gamma$ is equal to $0.4$ for equal mass clusters, and $t_{cross,ext}$ is the time necessary for a BH to cross the cluster in the presence of an external potential.

\begin{figure*}[htbp]
    \centering
    \includegraphics[width=\textwidth]{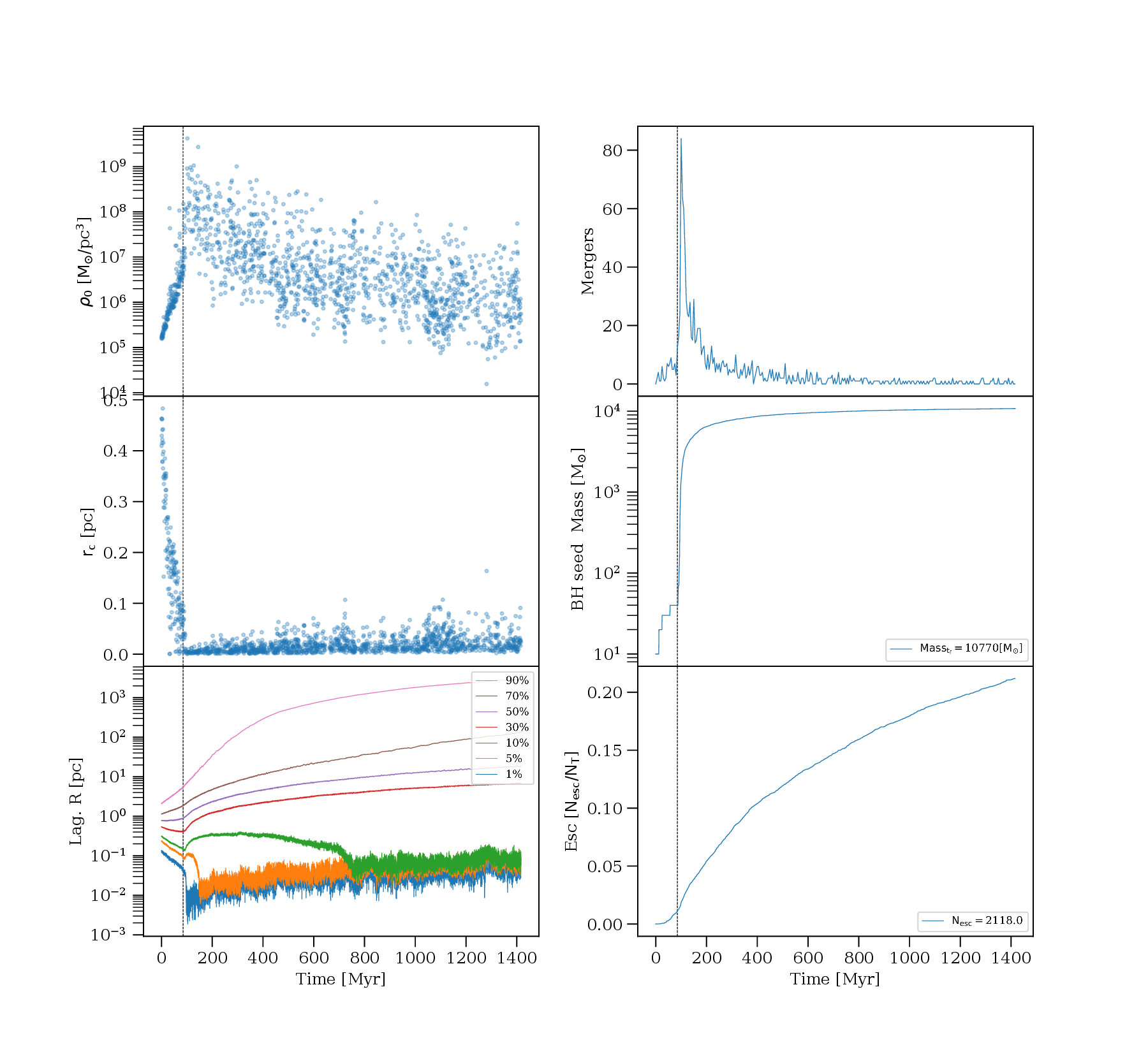}
    \caption{Evolution of the cluster in a simulation with speed of light c  = $  10^{3} \:  \mathrm{km/s}$  and an external potential of $\eta_{g} = 0.0$. The first two panels to the left present the results of our fit considering a King's model. In the top left panel, we display the density of the core, while in the left second panel, we illustrate the core radius of the cluster. In the bottom left panel we show the Lagrangian radius for mass fractions between $1\%$ and $90\%$ of the total mass of the cluster. The top right panel shows the mergers of BHs in the cluster in bins of $5 \: \mathrm{Myr}$, in the middle right  panel we show the growth of the mass of the most massive BH in the cluster. The bottom right  panel shows the accumulative ejections in the cluster. The vertical line  in the panels corresponds to the moment when the highest central density is reached.}
    \label{fig:2:c1_EP00}
\end{figure*}

In Fig.~(\ref{fig:2:c1_EP00}), we illustrate the evolution of the dark cluster without an external potential (i.e., $\eta_g = 0.0$) while considering a speed of light of $10^{3} \:  \mathrm{km/s}$. The crossing time of the cluster, assuming no external potential, is calculated to be $0.0482 \:  \mathrm{Myr}$. Additionally, the half-mass relaxation time as given by Eq. (\ref{eq:relaxation_time}) is $166.38 \:  t_{cross}$. The cluster reaches its highest density at $85.194 \: \mathrm{Myr}$ or, in terms of the half-mass relaxation time, at $10.61 \: t_{rh}$. The inner parts of the cluster as measured via the $10\%$ Lagrangian radius reach the highest density, $3.3 \times 10^{6} \: \mathrm{M_{\odot}/pc^{3}} $  at $0.144 \:  \mathrm{pc}$. 

In the left first panel we show the evolution of the core density and in the second panel we show the core radius evolution with time. The method used to obtain this results is explained in the next section. We note that before the core collapse the density of the core of the cluster has a steep increase reaching the maximum at $ \approx 10^{9}$ M$_{\odot}$/pc$^{3}$. At the same time the core radius reaches the minimum $\approx 0.01$ pc when the core collapse occurs (vertical line), after that the cluster begins to expand decreasing its density and increasing its core radius slowly. We can also note that the core density and core radius exhibit high dispersion in their trends probably because almost all the mass in the core is within the massive BH seed at the center of the cluster. In the bottom left  panel we show the Lagrangian radius, we observe that the $1\%$ Lagrangian radius post core collapse  experiences the motion of the central object because the central object has more than $1\%$ of the total mass of the cluster, the $5\%$ Lagrangian radius shows  a rebound, and approximately $50 \:  \mathrm{Myr}$ later, also shows a similar behavior like $1\%$ Lagrangian radius. A similar phenomenon occurs with the $10\%$ Lagrangian radius but with a longer delay in the collapse and subsequently the motion of the central object. Lagrangian radii greater than $10\%$ are affected by the expansion of the cluster. The middle right panel depicts the growth of the mass of the central object and the beginning of massive black hole formation. At the time when the highest density is reached, the growth becomes exponential, occurring in a short span of approximately $10 \: \mathrm{Myr}$, eventually reaching a mass of $10770 \: M_{\odot}$ by the end of the simulation.

In the third panel, we show the evolution of BH escapers from the cluster with a similar peak when the highest density is reached, resulting in a total mass loss of $21\%$ from the cluster. The fourth panel illustrates the evolution of mergers, with a peak of approximately $\approx 80$ mergers in 5~Myr. There is a second peak occurring approximately $\approx 50 \: \mathrm{Myr}$ later, with about $\approx 30$ mergers, coinciding with the contraction of the $5\%$ Lagrangian radius. 


\begin{figure*}[htbp]
    \centering
    \includegraphics[width=\textwidth]{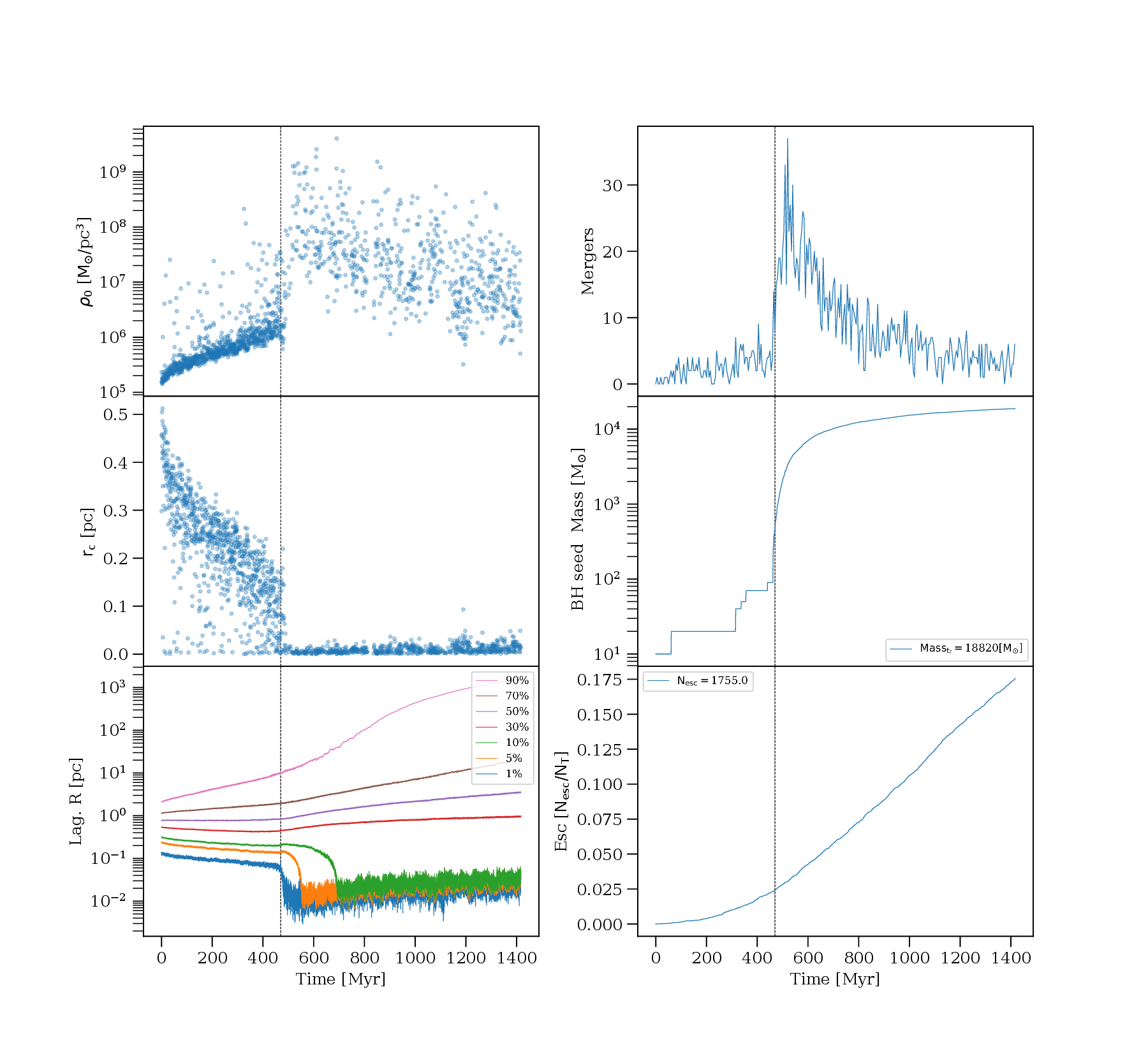}
    \caption{Evolution of the cluster in a simulation with speed of light c  = $  10^{3} \:  \mathrm{km/s}$  and an external potential of $\eta_{g} = 1.0$. The first two panels to the left present the results of our fit considering a King's model. In the top left panel, we display the density of the core, while in the left second panel, we illustrate the core radius of the cluster. In the bottom left panel we show the Lagrangian radius for mass fractions between $1\%$ and $90\%$ of the total mass of the cluster. The top right panel shows the mergers of BHs in the cluster in bins of $5 \: \mathrm{Myr}$, in the middle right  panel we show the growth of the mass of the most massive BH in the cluster. The bottom right  panel shows the accumulative ejections in the cluster. The vertical line  in the panels corresponds to the moment when the highest central density is reached.}
    \label{fig:3:c1_EP10}
\end{figure*}

In Fig.~(\ref{fig:3:c1_EP10}), we show the evolution of the cluster considering an external potential of $\eta_{g} = 1.0$ and a speed of light of $10^{3} \: \mathrm{km/s}$, The crossing time is $0.0241 \: \mathrm{Myr}$, and the half-mass relaxation time is $2662.149 \: t_{cross}$. The cluster experiences a high increase of the central density at $450 \: \mathrm{Myr}$ or, in terms of the half-mass relaxation time, $7.010 \: t_{rh}$. The density reached at the $10\%$ Lagrangian radius is $1.23 \times 10^{6} \: \mathrm{M_{\odot}/pc^{3}}$, with a radius of $0.2 \:  \mathrm{pc}$. The behavior of the cluster is remarkably similar to that of the cluster without an external potential, one of the differences is the delay in the contraction of the inner regions in the cluster and the lower contraction in the $10\%$ Lagrangian radius; 
also at the time of core collapse, the density and radius of the core exhibit even more abrupt changes considering a higher external potential with a steep increase in the density of the core and the core radius,
we have a higher merger rate and spread over a larger time interval, implying that the forming object becomes more massive. The amount of escapers is reduced by almost a $5\%$ compared to the cluster without external potential.  



\begin{figure*}[htbp]
    \centering
    \includegraphics[width=\textwidth]{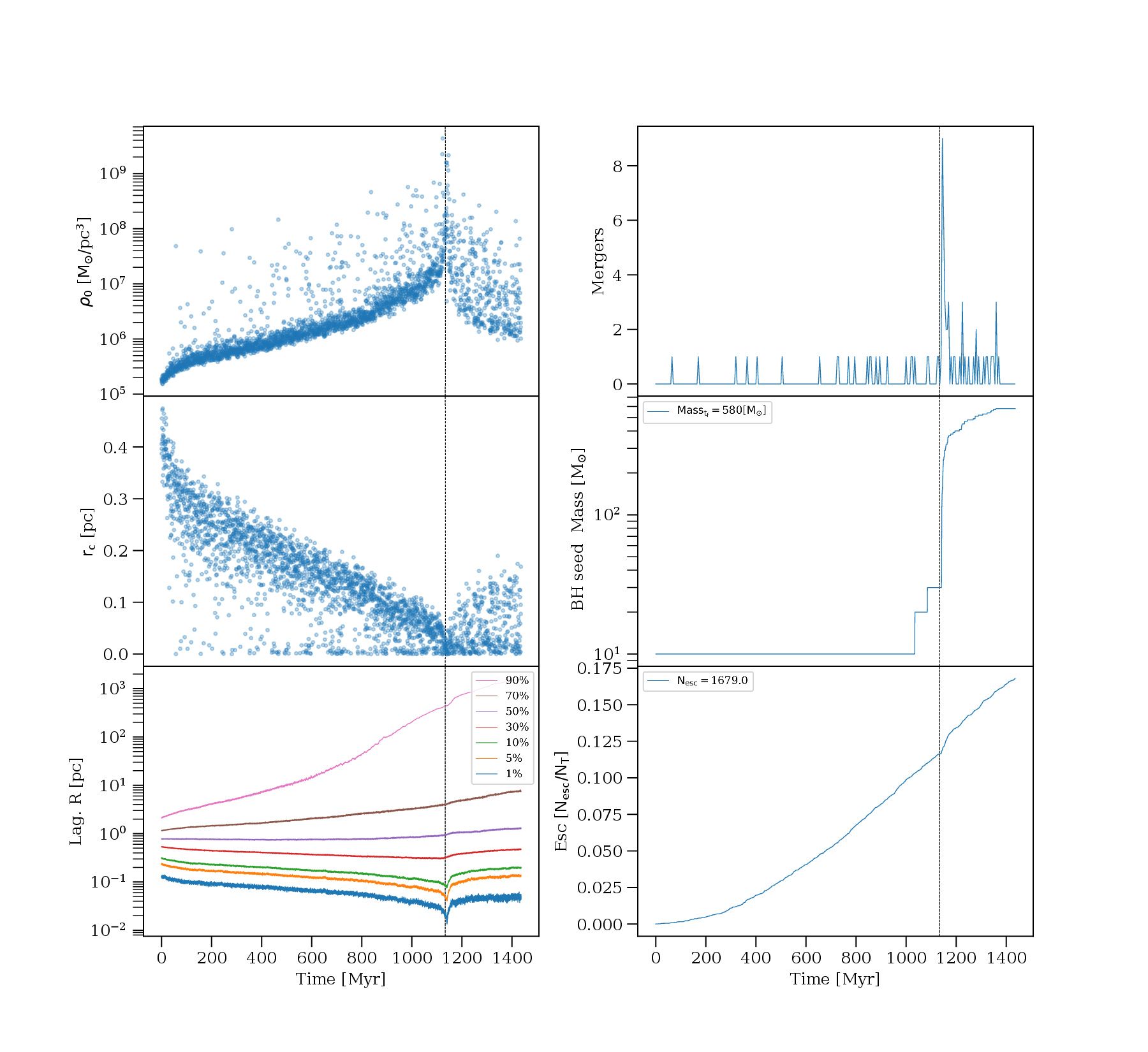}
    \caption{Evolution of the cluster in a simulation with speed of light c  = $ 3 \times 10^{4} \:  \mathrm{km/s}$  and an external potential of $\eta_{g} = 1.0$.  The first two panels to the left present the results of our fit considering a King's model. In the top left panel, we display the density of the core, while in the left second panel, we illustrate the core radius of the cluster. In the bottom left panel we show the Lagrangian radius for mass fractions between $1\%$ and $90\%$ of the total mass of the cluster. The top right panel shows the mergers of BHs in the cluster in bins of $5 \: \mathrm{Myr}$, in the middle right  panel we show the growth of the mass of the most massive BH in the cluster. The bottom right  panel shows the accumulative ejections in the cluster. The vertical line  in the panels corresponds to the moment when the highest central density is reached.}    
    \label{fig:4:c30_EP10}
\end{figure*}

Simulations with higher speed of light tend to decrease the phase of oscillations in the inner regions of the clusters. Additionally, they tend to delay the core collapse even when considering the same external potential. Moreover, the number of mergers decreases, resulting in lighter SMBH seeds. In Fig.~(\ref{fig:4:c30_EP10}), we illustrate the evolution of a BH cluster with an external potential of $\eta_{g} = 1.0$ and a speed of light of $3 \times 10^{4} \: \mathrm{km/s}$. The highest density we compute, considering the $10\%$ Lagrangian radius corresponding to a density peak of $1.92 \times 10^{7} \: M_{\odot}/\mathrm{pc}^{3}$, occurs at $1137 \: \mathrm{Myr}$ or, in terms of relaxation time, $17.77 \: t_{rh}$. As mentioned previously, there is a delay in the core collapse compared to clusters with the same external potential but a lower speed of light.   The core collapse appears to occur more smoothly compared to the case with a lower speed of light and is not as abrupt. Following the core collapse during the rebound process, the core density experiences a quick decrease. Additionally, the core radius undergoes a rapid increase, indicating that the rebound process is more prominent when the cluster has a higher speed of light. The contractions  occur at the same time in the inner regions of the cluster, and prominent motion of the central object is  only observed in the $1\%$ Lagrangian radius. The number of escapers in the cluster is similar to the other clusters with the same external potential, with only $17\%$ of BHs escaping.



\subsection{Time dependence of core collapse}

The most common way to determine whether a core collapse occurs in an N-body system is via the evolution of the core radius and the core density.  An additional possibility involves considering the binding energy of binaries within the core of the cluster. During core collapse, binaries harden through three-body interactions until they possess enough energy for the core to bounce \citep{Fujii_2014}. In the analysis presented here, we focus on the first possibility, the bounce of density and radius, as the density peak is significant enough to be visually observed. This is well-justified as our model involves equal mass BHs so implying low values of $f_{max}=m_{max}/<m>$, while core collapse becomes more ambiguous for larger values of  $f_{max}$ \citep{Fujii_2014}.

We estimate the core radius and core density by fitting a density profile according to King's model {\citep{king_1962}},
\begin{equation} \label{eq:King_density}
    \rho_{King} = \rho_{0} \left( 1 + \left(\frac{r}{r_c} \right)^{2} \right)^{-3/2}.
\end{equation}

This fitting is considering the cluster without the central massive object at the moment of core collapse. The massive object is not included as its increase in mass via collisions may contribute to an increase in central mass density, though our interest here is to determine whether there is a contraction of the central core. We determine the time of core collapse via the evolution of the peak density of the core.

As we increase the external potential, one of the significant differences is the time it takes for the contraction of the inner regions  to occur. As observed in the previous sections, there is a difference of more than $1 \: \mathrm{Gyr}$ between the cluster without a gas potential and the one with an equal mass fraction of gas and BHs, considering a speed of light of $3 \times 10^{4} \: \mathrm{km/s}$. On the other hand, at a lower speed of light of $10^{3} \:  \mathrm{km/s}$, this difference in the time delay between the simulations with the highest and the lowest external potential corresponds to only about $450 \: \mathrm{Myr}$. For simplicity and to adopt a uniform approach between the simulations, we employ the $10\%$ Lagrangian radius to determine the time of maximum contraction corresponding to the central density peak.

\begin{figure}[htbp]
    \centering
    \includegraphics[width=0.5\textwidth]{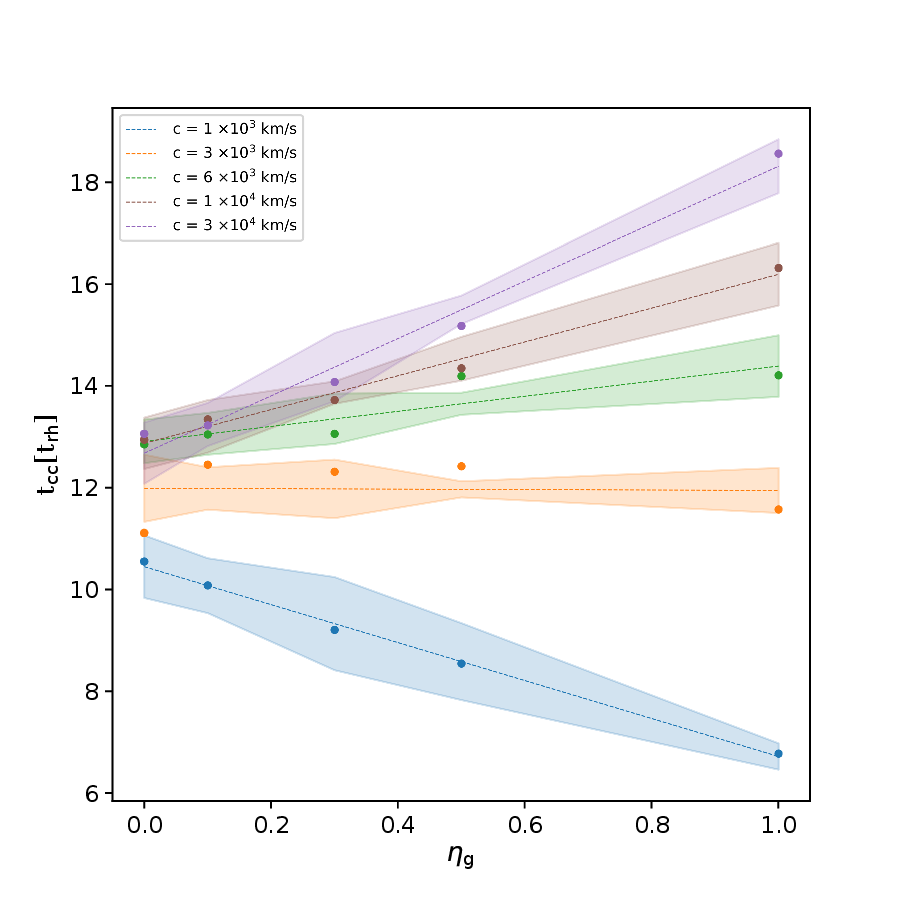
}
    \caption{We show the core collapse time relative to the half-mass relaxation time as a function of the gas mass fraction of the cluster, denoted as $\eta_{g}$. Each curve represents a different value of the speed of light $c$. The shadow zone is the error computed via the standard deviation from simulations with different initial conditions.}    
    \label{fig:5:tcc_trh}
\end{figure}

In  Fig.~(\ref{fig:5:tcc_trh}) we show the time of the core collapse in the cluster in relaxation time given by Eq.~(\ref{eq:relaxation_time}) as a function of the external potential ($\eta_{g}$) at different speed of light. The maximum central contraction is reached within 6-20 half-mass relaxation times, as evident in Fig.~(\ref{fig:5:tcc_trh}). Assuming that core collapse is proportional to the relaxation time \citep{Spitzer1987}, we can infer that the time of contraction of the inner regions is proportional to $t_{cc} \propto (1+\eta_{g})^{4}t_{cross}$, so the time of core contraction tends to be higher when the external potential increases \citep{Reinoso2020}. The linear trend suggests that clusters are more affected by gravitational radiation if the speed of light is reduced, thus making them more relativistic. In simulations where the speed of light is particularly low, i.e. less than $3\times10^3$~km/s, it is conceivable that the contraction time is even enhanced due to efficient gravitational wave emission. This is supported by simulations with speed of light $\leq 3 \times 10^{3} \: \mathrm{km/s}$ where the rms is larger than $1\%$ of the value of the speed of light used in the simulation, implying that the BH cluster is in a relativistic state \citep{Kupi_2006}. For instances in the cluster with $c=10^{3} \: \mathrm{km/s}$,  the rms velocity is higher than $1\%$ of the speed of light  that we consider in the simulation IDs 1-5. Furthermore, the relativistic state is more prolonged for higher external potentials, as the rms speed increases with the external potential, affecting the cluster via strong relativistic effects leading to the dissipation of kinetic energy into gravitational waves. For speeds of light exceeding $c = 3 \times 10^3 \: \mathrm{km/s}$, we observe that the rms speed is slightly below $1\%$ of the speed of light {considered in this simulation} ($30$~km/s), but it is very close. Consequently, we might expect that gravitational radiation is not exceptionally strong, but it is still sufficient to reduce the time for contraction of the cluster. On the other hand,  the external potential increases the core collapse timescale. This is evident when examining the orange curve in Fig.~(\ref{fig:5:tcc_trh}). However, for higher speeds of light, gravitational radiation is not strong enough, leading to a delay in the contraction of the inner region.



\subsection{Binary population}

\begin{figure}[htbp]
    \centering
    \includegraphics[width=0.5\textwidth]{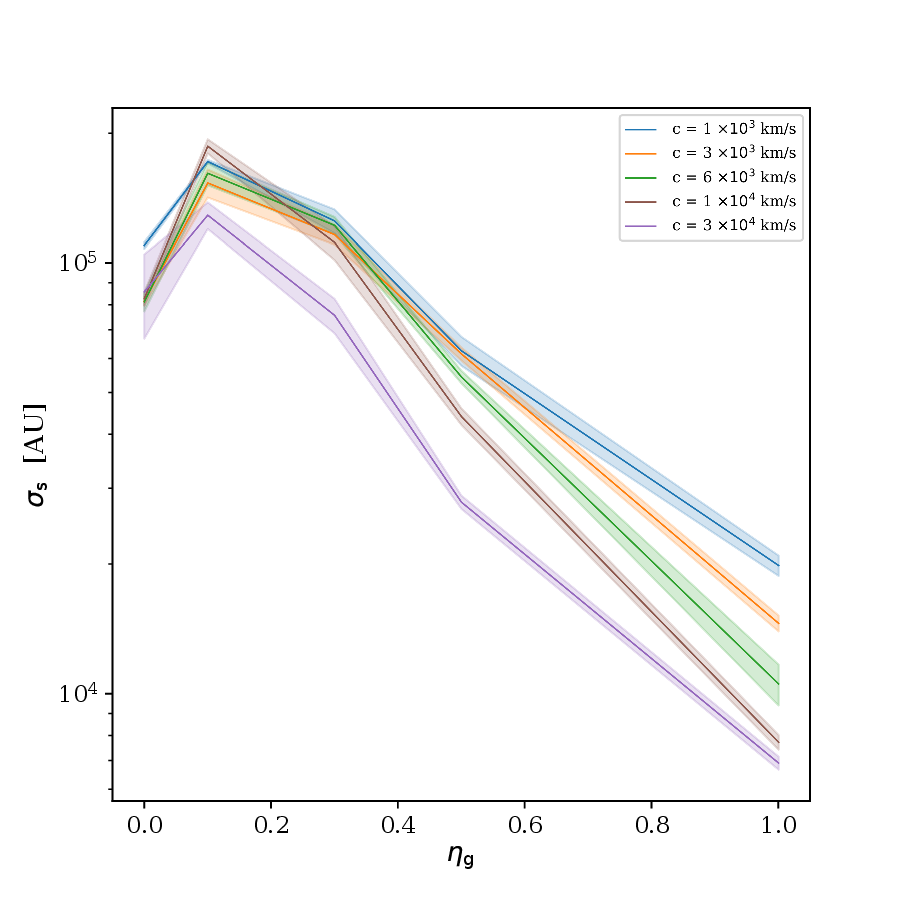}
    \caption{We show the root-mean square dispersion of the semi-mayor axes of the binaries in the cluster as a function of the external potential ($\eta_{g}$). The different colors correspond to different speeds of light. The shadow regions correspond to the error associated to the random initial conditions.}    
    \label{fig:6:Esc_semi}
\end{figure}

Our simulations indicate that with respect to both the binary population and the mass of the cluster, the population of binary systems decreases when the cluster experiences a deeper external potential. This trend is primarily attributed to the disruption of soft binaries resulting from the increase in the velocity dispersion within the cluster. In dense star clusters, binaries are influenced by two-body encounters, leading to a drift due to mass segregation. This is primarily because binaries possess a larger mass relative to single stars. In denser regions, the semi-major axis of binary systems tends to decrease over time, which leads to an increase in their hardness or their disruption via encounters with single BHs. To provide a clearer view of the trends in the semi-major axis at different external potentials, we calculated the standard deviation of the semi-major axis of all binaries that are formed in the cluster via third-body BH interaction, finding an increase in the semi-major axis up to $\eta_{g} = 0.1$ see Fig.~(\ref{fig:6:Esc_semi}). As the external potential increases, it becomes evident that binaries tend to become more tightly bound, resulting in a significant reduction in the spread of the semi-major axis, nearly by one magnitude, when $\eta_{g} = 1.0$. This trend is similar to the behavior  of the number of  escapers in the cluster, and  could be a result of weak interactions increasing the kinetic energy enough to eventually escape from the cluster given the higher cross section of the binaries of cluster with higher external potential. For larger values of the speed of light, the semi-major axis tend to slightly decrease, as the evolution of the binaries via gravitational wave emission is decelerated.



\subsection{IMBH seed formation }


\begin{figure*}[htbp]
    \centering
    \includegraphics[width=1\textwidth]{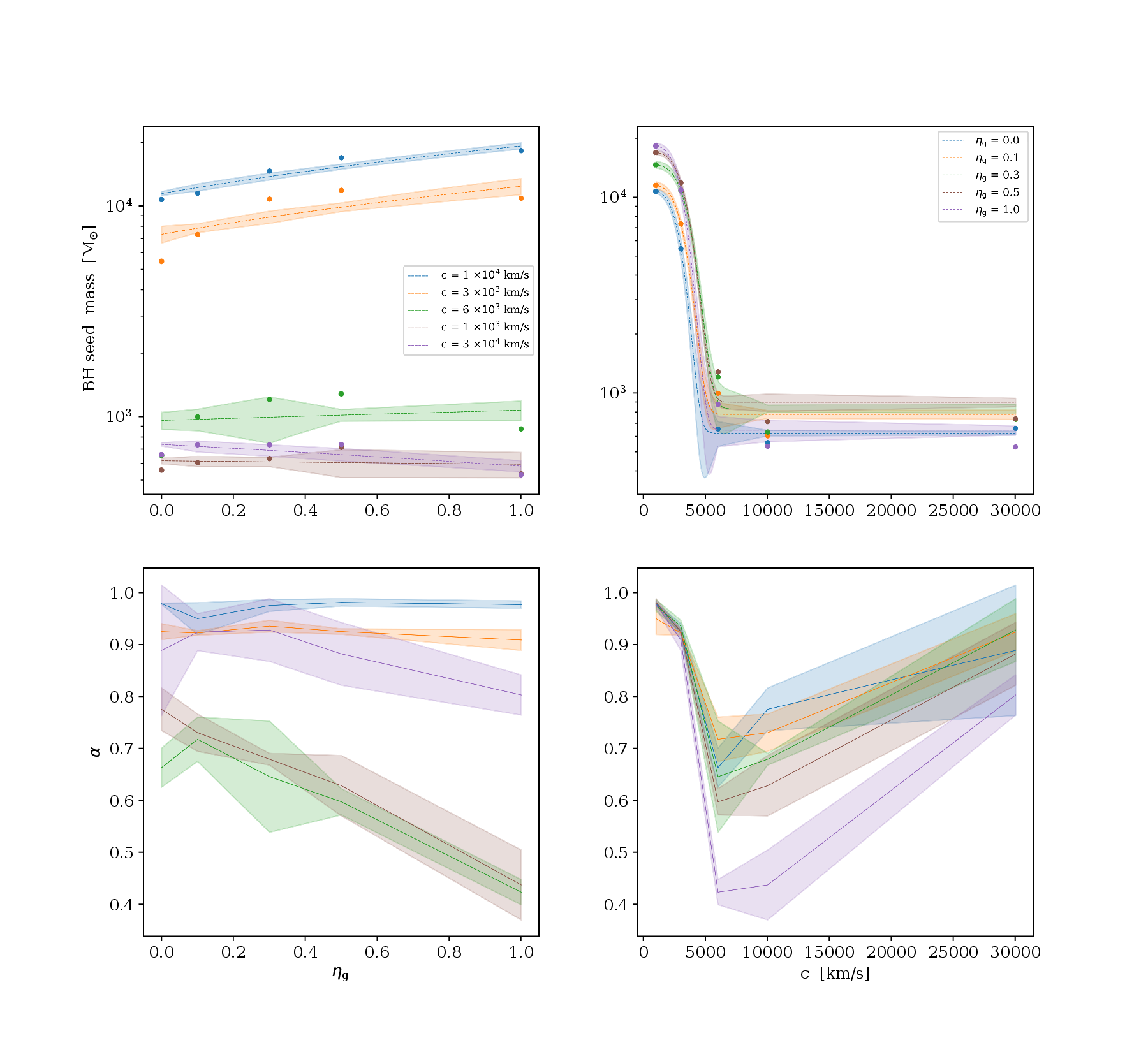}
    \caption{In the top panels, we display the mass of the most massive black hole (BH) in the cluster at the end of the simulation as a function of the external potential $\eta_{g}$ (left) and as a function of the speed of light (right). In the bottom panels, we provide the parameter $\alpha$ as a function of the external potential  $\eta_{g}$ (left) and as a function of the speed of light  (right). This parameter is defined as the ratio of total  mergers in the cluster and  the mass of the of the massive object, indicating how many merging objects contribute to the formation of the central object.}
    \label{fig:7:fit_mass}
\end{figure*}

The mass of the most massive object formed in the simulations is provided in Fig.~(\ref{fig:7:fit_mass}) as a function of the $\eta_g$ (left panel) and as a function of the speed of light $c$ (right panel). For speeds of light of $6\times10^3$~km/s or higher, the mass of the most massive object is essentially independent of $\eta_g$ and in the range of $500-1500$~M$_\odot$, 

At least for $\eta_{g} < 1.0$, we can observe a slight decrease in the trend of the mass in the central object. This occurs as the time of contraction increases with higher external potential as we can see  Fig.~(\ref{fig:5:tcc_trh}), and the central object has lower time post core collapse to increase its mass by mergers.

For lower speeds of light, the formation of the massive object has been considerably enhanced by stimulated gravitational wave emission, leading to masses of the most massive object of the order $10^4$~M$_\odot$. A higher vale of $\eta_g$ further stimulates gravitational wave emission and enhances this effect. Looking at the mass of the most massive object as a function of $c$, we note that from the right panel of Fig.~(\ref{fig:7:fit_mass}) that the relation considerably flattens for speeds of light of $6\times10^3$~km/s or higher. 

As observed in Fig.~(\ref{fig:2:c1_EP00}) and Fig.~(\ref{fig:4:c30_EP10}), mergers of black holes can occur independently of the core contraction event. This suggests that the conditions for these mergers are not exclusively confined to the core contraction phase. Mergers can also occur after this contraction event due to the high density in the inner regions of the cluster. The presence of numerous binaries surrounding the central region and recently formed massive objects may enhance BH binary mergers via the Kozai-Lidov mechanism \citep{Aarseth_2007,Sedda_2020}. This mechanism involves the attainment of large eccentricities through third-body secular perturbations, and finally feeding the central object. Other speculative mechanism that could enhance the eccentricity of binaries in dense clusters is via perturbation of single objects passing near  the binary system driving the eccentricity to 1 \citep{Reinoso_nathan_2022},

Additionally, the external potential has a significant impact on the binary population by reducing the number of binaries available for mergers. This reduction of binaries could affect the mass of the central massive black hole. Moreover, the external potential also delays the timing of core contraction. However, it is important to note that the density and velocity dispersion within the cluster are essential factors for the formation of "hard" binaries, which are more likely to merge due to gravitational radiation.

In Fig.~(\ref{fig:7:fit_mass}), we further report the parameter $\alpha$, which correspond to the ratio 
between the number of mergers and the mass of the most massive object, both as a function of $\eta_g$ and $c$. For speeds of light of $6\times10^3$~km/s or less, the ratio $\alpha$ has only a weak dependence on $\eta_g$ and is in the range of $0.85-1$. In the simulations with higher speeds of light, $\alpha$ decreases with $\eta_g$, from values around $0.75$ in the absence of an external potential to $\alpha\sim0.45$ for $\eta_g=1$. When we considering the dependence on the speed of light we find a weak dependence on $\eta_g$ for speeds of light below $3\times10^3$~km/s, as the efficient gravitational wave emission leads to a rapid contraction of the system. When $c$ increases, there is more time for mergers among stellar mass BHs to occur, with a more significant spread and dependence on $\eta_g$. With further increasing values of $c$, $\alpha$ increases again for all values of $\eta_g$ as larger values of $c$ also suggest a longer timescale for the merger of stellar mass BHs, while the timescale for the central collapse remains very similar.

\begin{figure}[htbp]
    \centering
    \includegraphics[width=0.5\textwidth]{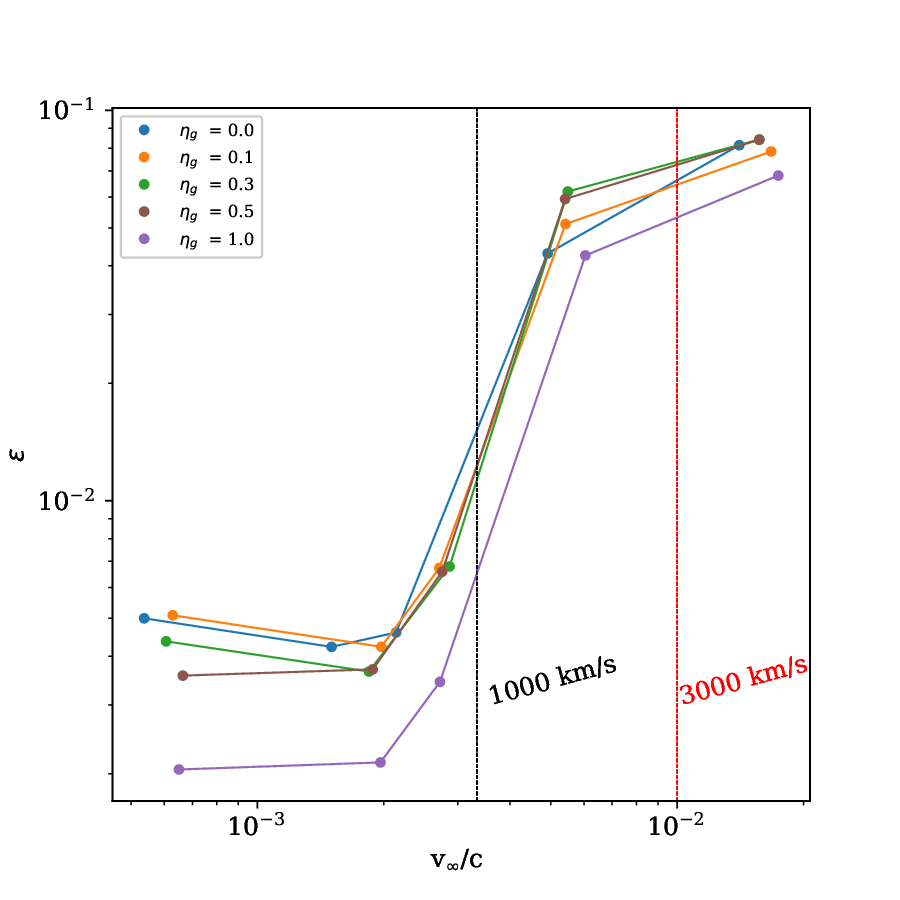}
    \caption{We present the BH formation efficiency of the clusters defined as the mass of the most massive BH divided by the total BH mass  of the cluster, as a function of the ratio between the root mean square (rms) velocity at the time of the core collapse  and the speed of light ($c$) employed in the simulations. Different colors are used to denote varying external potentials $\eta_{g}$. The vertical lines mark  velocity ratios assuming the real value of the speed of light for clusters with an rms velocity of $1000$ $\mathrm{km/s}$  and $3000$ $\mathrm{km/s}$. }    
    \label{fig:8:efficiency}
\end{figure}

The efficiency for the formation of the most massive object, defined as its final mass divided by the cluster mass in BHs, is provided in Fig.~(\ref{fig:8:efficiency}) as a function of the ratio of the root mean square velocity $v_\infty$ at the time of core collapse divided by the speed of light adopted in the simulation. As a result, we find a clear relation, where the efficiencies are in the range of $0.001-0.004$ for values $v_\infty/c<0.0025$. The efficiencies increase to $0.04-0.07$ for $v_\infty/c\sim0.005$ and a further more moderate increase occurs for $v_\infty/c\sim0.015$, reaching efficiencies of $0.06-0.08$. The dependence on the external potential is however non-trivial; we saw already above that the mass of the most massive object can either increase with $\eta_g$, stay constant or slightly decrease, leading here to a complex relation as a function of $v_\infty/c$. Fig.~(\ref{fig:8:efficiency}) nonetheless allows to pursue a tentative extrapolation towards real systems, assuming black hole clusters with velocity dispersion of $1000$~km/s and $3000$~km/s together with the physical speed of light. As indicated within the figure, the latter corresponds to efficiencies of $0.004-0.05$ and $0.05-0.07$, respectively.

\subsection{Extrapolation to NSC}


\begin{figure*}[htbp]
    \centering
    \includegraphics[width=0.8\textwidth]{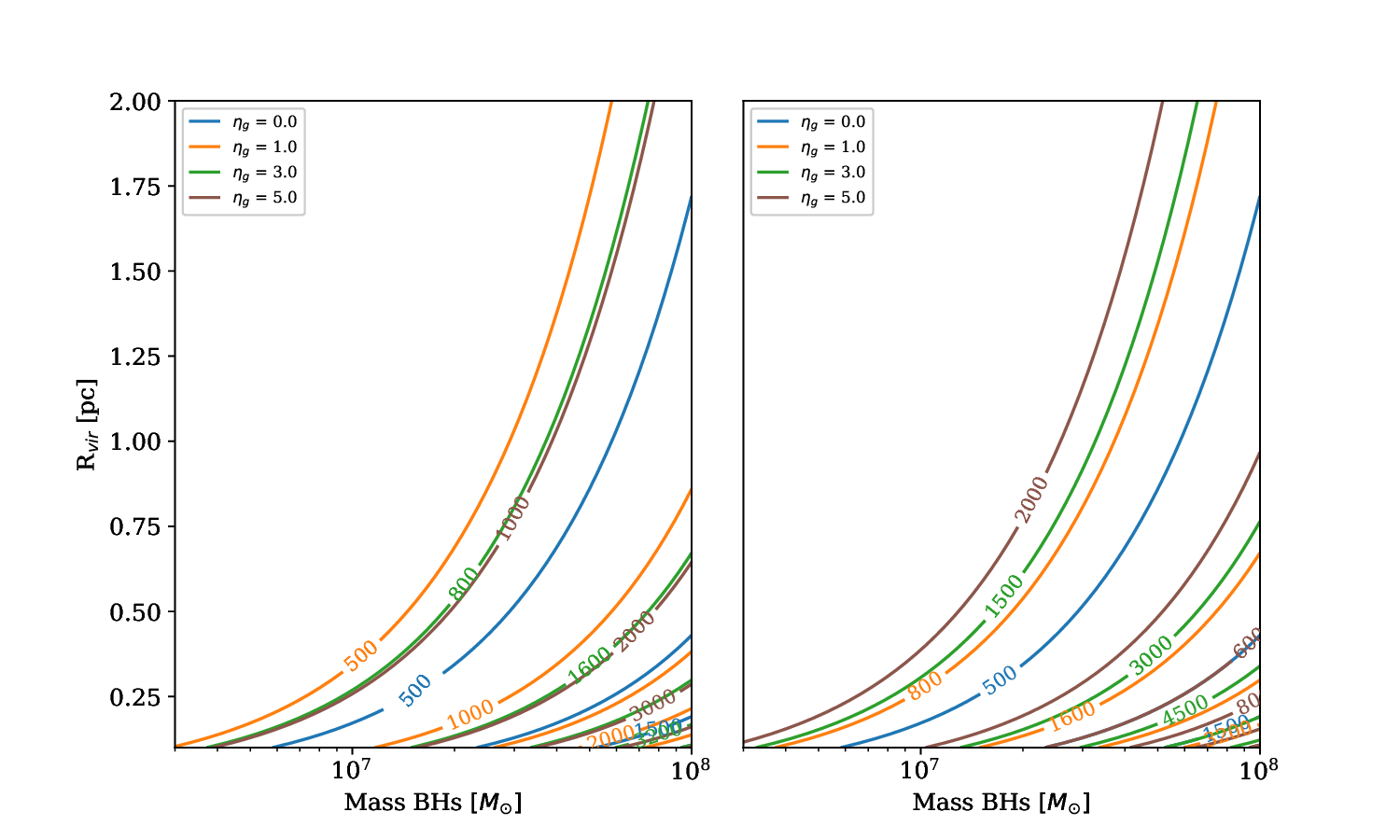}
    \caption{In the left panel, we show the rms contour lines of the velocity calculated via Eq.~\ref{eq:rms_vir}, providing its dependency on the virial radius and the total BH mass in the cluster. The contours illustrate the velocities at specific radii and masses. The right bottom panel provides countour lines of the velocity of the cluster as defined by Eq.~\ref{eq:rms_contraction}, assuming an additional contraction as calculated by \citet{Kroupa2020}. The different colors indicate various values of $\eta_{g}$.}    
    \label{fig:9:RMS}
\end{figure*}

\begin{figure*}[htbp]
    \centering
    \includegraphics[width=1\textwidth]{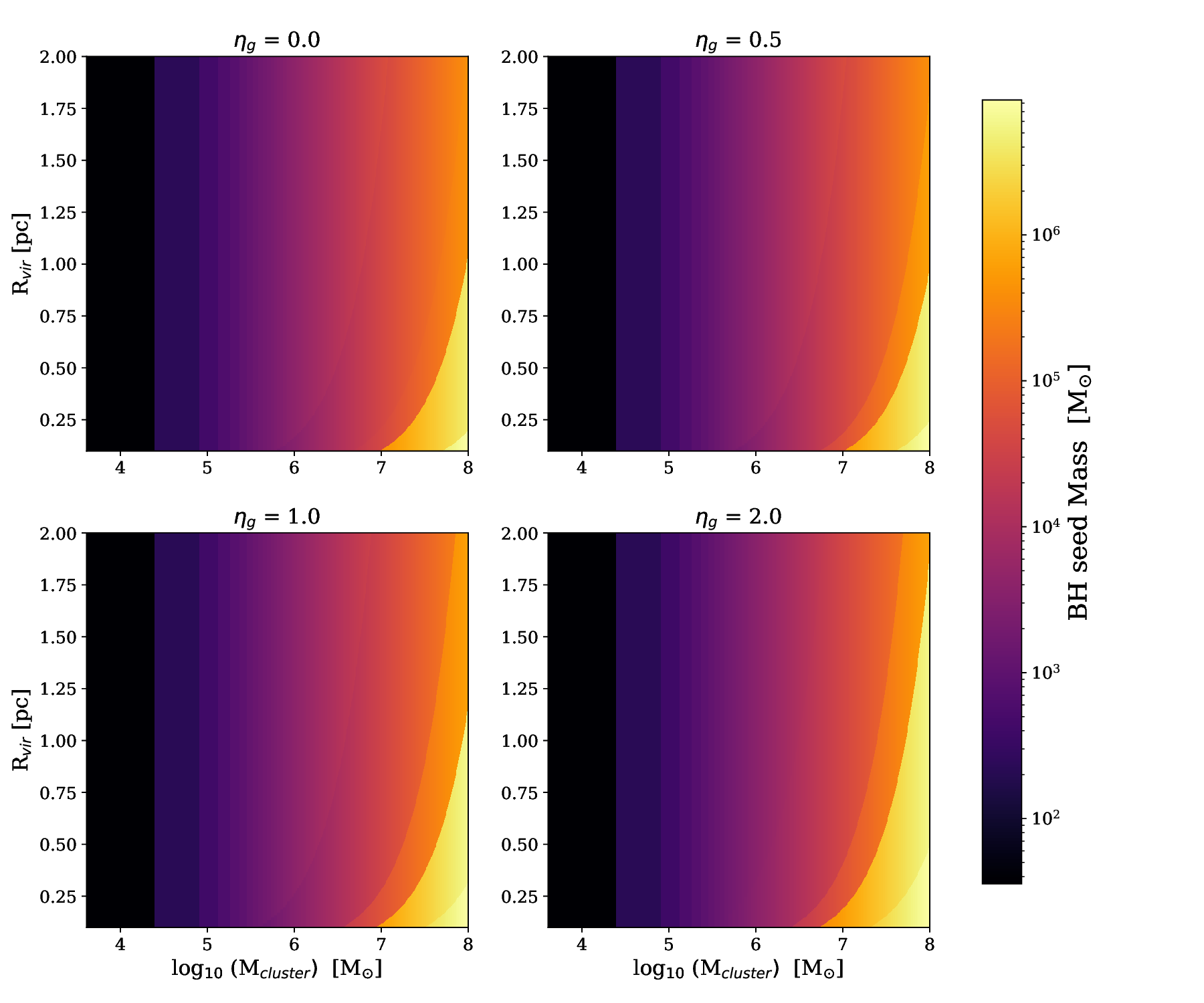}
    \caption{We provide the mass of the most massive  black hole estimated via the root mean square (rms) velocities calculated using Eq.~\ref{eq:rms_vir} and their corresponding efficiency depicted in Fig.~\ref{fig:8:efficiency}. The clusters are within a range of virial radii from $0.1$ to $2.0 \: \mathrm{pc}$ and masses from $10^{4}$ - $10^{8} \: \mathrm{M_{\odot}}$. The color represents the mass of the BH that forms. Each panel corresponds to a different value of the external potential.}    
    \label{fig:10:IMBH_vir}
\end{figure*}

\begin{figure*}[htbp]
    \centering
    \includegraphics[width=1\textwidth]{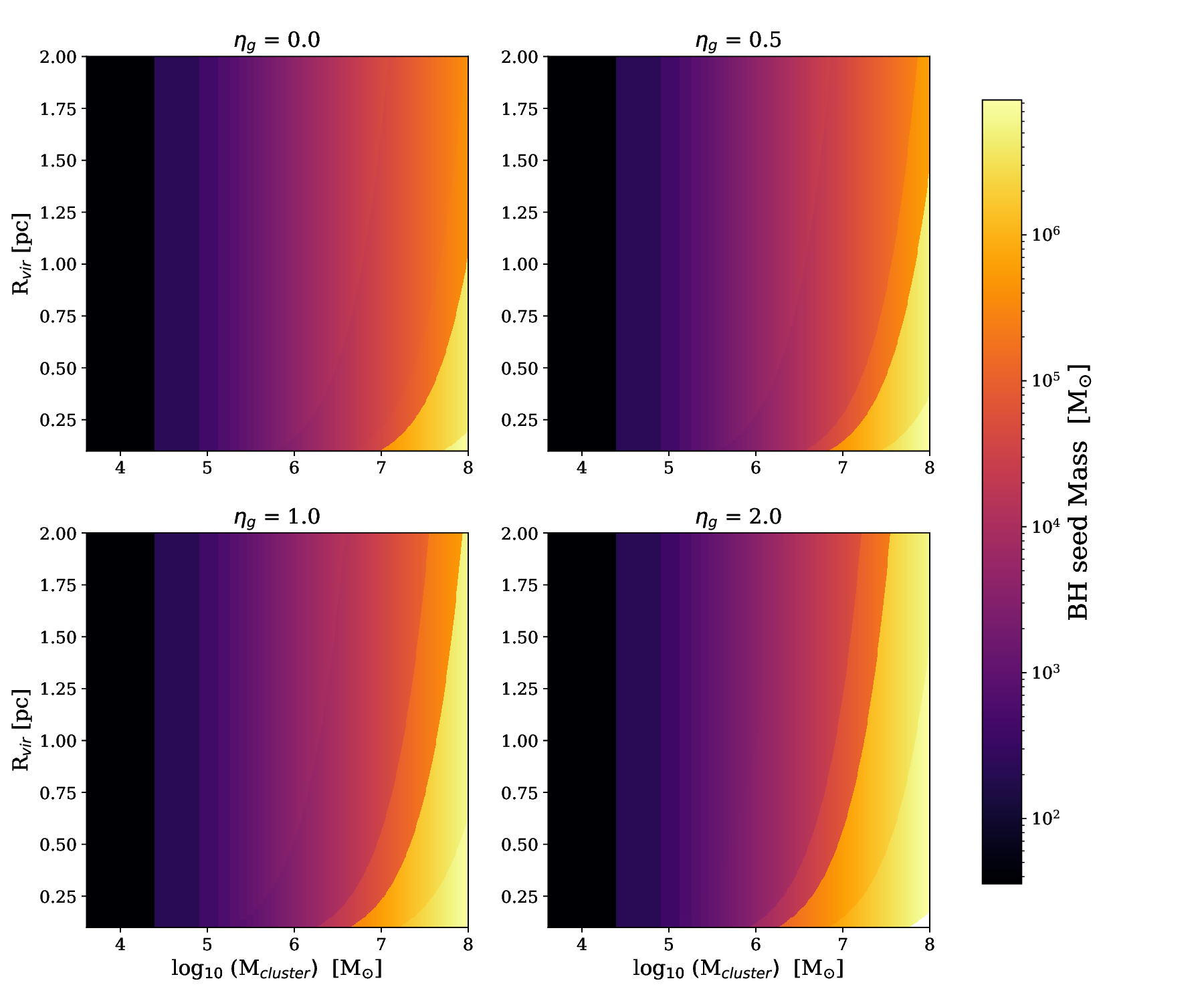}
    \caption{We provide the mass of the most massive black hole (BHs) that can form calculating the root mean square (rms) velocity  using Eq.~\ref{eq:rms_contraction} from \citet{Kroupa2020} and and the corresponding efficiency depicted in Fig.~\ref{fig:8:efficiency}. The clusters are within a range of virial radii from $0.1$ - $2.0 \: \mathrm{pc}$ and masses from $10^{4}$ - $10^{8} \: \mathrm{M_{\odot}}$. The color represents the mass of the BH that forms. Each panel corresponds to different external potentials.}    
    \label{fig:10:IMBH_K}
\end{figure*}

After the tentative extrapolation of the efficiencies to real astrophysical systems in the previous subsection, we here aim to assess more quantitatively the masses of the most massive objects that could be formed via mergers in dense black hole clusters as a function of the physical conditions. We show in the previous subsection that a central parameter is the ratio $v_\infty/c$, as it allows to characterize the relativistic state of the cluster and provides a good indication of the efficiency for the formation of a very massive black hole.

As a first step, we thus aim to infer the rms velocities that can be expected in different clusters as a function of their mass and radius. For this purpose, we consider and distinguish two different cases: The first scenario assumes that the black holes in the cluster are in equipartition with their self-gravity and the external potential, as intrinsically assumed in the simulations we presented here. In this case, we have

\begin{equation}\label{eq:rms_vir}
    v_{\infty} \approx  \sqrt{ 0.4 \frac{G M_{BHs} (1+ \eta_{g})}{R_{vir}}}.
\end{equation}
$v_{\infty}$ however does not adequately account for the contraction of the black hole cluster due to gas accretion. Therefore, the cluster is less dense than considered in the model proposed by \citet{Davies2011}, where gas accretion affects the density of the cluster  and subsequently influences the root mean square velocity. A more accurate velocity dispersion for a cluster affected by gas accretion was derived by \citep{Kroupa2020} as

\begin{equation}\label{eq:rms_contraction}
    \sigma = \left( f_{v}   \frac{G M_{BHs}(1+\eta_{g})^{2}}{ R_{vir} } \right)^{1/2}, 
\end{equation}
where $M_{BHs}$ is the mass of BH in the cluster, $R_{vir}$ is the virial radius and $f_{v} \approx 1$ is a dimensionless factor that covers a departure from the virial equilibrium or a particular shape of the potential well. The corresponding velocity dispersion is given as a function of BH mass in the cluster and the cluster virial radius in Fig.~(\ref{fig:9:RMS}) both for the case where they are calculated via Eq.~(\ref{eq:rms_vir}) and Eq.~(\ref{eq:rms_contraction}). The behaviour in both cases is similar but the velocity dispersions is somewhat enhanced in the second case. Particularly high velocity dispersion occur when both the total mass in the cluster is high and the virial radius of the cluster is small. To obtain velocity dispersions of the order $3000$~km/s, implying relativistic clusters, it seems likely that cluster masses of $10^7$~M$_\odot$ or higher are required.

The mass of the most massive object that forms in the two cases is finally estimated employing the efficiencies from Fig.~(\ref{fig:8:efficiency}). In the first case where the velocity is calculated assuming virial equilibrium  Eq.~(\ref{eq:rms_vir}), the resulting black hole masses are provided as a function of the cluster mass and radius in Fig.~(\ref{fig:10:IMBH_vir}), where the different panels correspond to different values of $\eta_g$. In case of black hole clusters with masses of $10^5-10^6$~M$_\odot$, our model suggests that only black hole seeds of moderate masses can be formed, roughly of order $10^3$~M$_\odot$. Black hole seeds of $10^4$~M$_\odot$ require clusters with at least $\sim10^{7}$~M$_\odot$ that still needs to be sufficiently compact with a virial radius of less than a parsec. The presence of an external potential potentially makes this more feasible, as it tends to increase the range of virial radii for which a massive object of significant mass can be produced for otherwise equal cluster parameters.  

We also provide the corresponding estimates for the scenario provided by \citet{Kroupa2020}, with the estimated black hole masses given in Fig.~(\ref{fig:10:IMBH_K}) for different values of $\eta_g$. The overall behavior is similar, but for a stronger sensitive to the velocity dispersion of the cluster; however, We note a shift by about half an order of magnitude lower in the required mass of the cluster for forming a seed black hole of a certain stellar mass, which is dependent on the external potential. The decrease of the cluster radius and  increase its density as a result of the gas inflow is thus potentially relevant and may lead to the formation of more massive black hole seeds under otherwise similar conditions.

\section{Discussions}  \label{section:Discussions}

This investigation provides a clear insight into the formation of an IMBH in the dark core of a NSC. In a simplified model, we consider a cluster with equal mass BHs distributed according to a Plummer distribution. In our simulations, we form two sets of BH seeds with approximately $10^4$~M$_\odot$ for clusters in a relativistic state and approximately $10^3$~M$_\odot$ for other clusters.

However, our models are affected by different assumptions. For instance, assuming equal-mass BHs in the cluster impacts mass segregation, leading to time scales for cluster evolution higher than in reality. Particularly, in clusters with a realistic stellar mass function, we have  $t_{cc} \backsimeq 0.2t_{rh}$ \citep{PortegiesZwart_2002}. Additionally, we neglect gas accretion onto the BHs, which further affects the time scale evolution, including the relaxation time \citep{leigh_2013} and the mass distribution of the BHs in the cluster. Furthermore, gravitational recoil caused by gravitational waves was not included in the simulations presented here. Studies such as \citet{Fragione_2020} provide a more extensive analysis of mergers and escapers considering recoils, with velocities up to a few thousand $\mathrm{km/s}$, where the typical mass of an ejected massive BH is 400-500 $M_{\odot}$. They also explore how the mass and density of the NSC influences the retention of massive BHs and the formation of binaries, where the massive NSCs can more easily retain massive BHs but the formation of binaries requires longer time scales. Dense NSCs  can both retain massive BHs and have a higher efficiency in forming binaries that merge through GW emission.

Regarding future work on the formation of an IMBH in a realistic NSC, \citet{Kroupa2020} demonstrate that for a high mass ratio of gas, $ \eta_{g} > 5.78 $, the cluster tends to expand for dark core masses  $< 10^{7} M_{\odot}$. However, for a mass of the dark core $ > 10^{7} \: \mathrm{M_{\odot}}$, the cluster is already initially in a relativistic state. To form an IMBH, we could consider massive dark cores with either massive black holes $ > 10 \:  \mathrm{M_{\odot}}$ or a higher number of BHs in the cluster. But this is not enough to reach the core collapse with the methodology that we use so far, as the  relaxation time scales as $\propto (1+\eta_g)^{4}$, so for values of $\eta_g$ higher than  $2.0 $  the core collapse requires more than $1.4 \: \mathrm{Gyr}$. We further note that an initial mass distribution of the BHs  reduces the time of the core collapse. 

Of course, this is a recent investigation and our knowledge on the mass distribution of stellar mass BHs is still limited, and no model can fully reproduce the distribution of observed total masses. Nevertheless, the observations lie within the distribution of mass in the 1$\sigma$ band of the $M_{max} = 50 \: \mathrm{M_{\odot}}$, $\alpha = 2.35$ model \citep{Perna_2019}. which could be employed in follow-up calculations in the future. In future projects it will also  be important to understand resonant relaxation (or Kozai) effects, which could significantly increase the rate of inspiral and their relation with the $\mathcal{PN} 1$ and $\mathcal{PN} 2$, affecting the precession and the impact of the number of captures \citep{Hopman_2006}. Finally, the consideration of radiation recoil will give us a better understanding of the evolution and the formation of IMBHs.


\section{Conclusions}  \label{section:Conclusions}
In this paper, we have explored the formation of seed black holes in dense black hole clusters embedded in an external potential, with the goal of exploring the hypothesis of \citet{Davies2011} that relevant gas inflows into such compact clusters will significantly increase the velocity dispersion and help to make the timescale for gravitational wave emissions more relevant compared to the timescale for ejections via 2+1 encounters, thereby favoring the formation of a central massive object.

As the simulations of massive black hole clusters incorporating post-Newtonian corrections with the real value of the speed of light are still computationally unfeasible, we have treated the speed of light as a free parameter to explore how the results of our simulations depend on the speed of light and more specifically also on the ratio of the velocity dispersion of the cluster divided by the speed of light. The latter allows us to test and explore the dependence on this parameter including an extrapolation towards the parameters of real physical systems. We focused on black hole clusters with $10^4$ stellar mass black holes with each of them having $10$~M$_\odot$, a virial radius of $1$~pc, ratios of gas mass to mass in BHs ranging from $0$ to $1$ as well as values of the speed of light ranging from $1000$~km/s up to $3\times10^4$~km/s. 

For values of the speed of light of $3000$~km/s or less, we found gravitational wave emission to be strongly enhanced, increasing even the contraction time of the cluster and favoring the formation of very massive objects of $10^4$~M$_\odot$ or more. For larger values of the speed of light, the sensitivity to this parameter significantly decreased as the timescale for gravitational wave emission was significantly enhanced, leading to typical seed masses in the range of $500-1500$~M$_\odot$. Particularly, when considering the ratio of cluster rms velocity $v_\infty$ divided by the speed of light $c$, the latter provided a good relation with the efficiency for the formation of the most massive object, which we defined as the mass of the most massive object divided by the total mass in BHs. The latter has ranged from $0.001-0.004$ for very low values of $v_\infty/c$ to values in the range of $0.05-0.08$ for $v_\infty/c\sim0.005-0.015$. 

When extrapolating to real astrophysical systems, we found that black holes clusters with masses in the range of $10^5-10^6$~M$_\odot$, this scenario may only be able to provide seed black holes of $\sim10^3$~M$_\odot$. In clusters of $10^7$~M$_\odot$ that are more compact than a parsec, the formation of seed masses with $\sim10^4$~M$_\odot$ is conceivable. The presence of an external potential allows the formation of such objects also in clusters of moderately increased size. If we adopt the formula provided by \citet{Kroupa2020} for the increase of the velocity dispersion as a result of the contraction due to the gas inflow, we further find that the required masses of the black hole clusters decrease by roughly half an order of magnitude.

{The astrophysical implications of the black holes formed via this mechanism will depend on their capability for subsequent growth. Even for accretion at a few percent of the Eddington rate, a strong radiatively driven wind could self-regulate black hole growth. The physics of such radiatively driven winds has been explored e.g. by \citet{Thorne1981} and \citet{Yamamoto2021}, which might be complemented by radiation driven disk winds as well \citep{Margherita2019}. Observational evidence suggests that feedback via radiation-driven winds was more frequent and stronger in the early Universe \citep{Bischetti2022} and may have been driven due to the opacity from dust grains \citep{Ishibashi2019}. Particularly winds from hot accretion flows were recently shown to be able to reach large scales \citep{Cui2020}, potentially expelling large fractions of the gas in the host galaxy \citep{Brennan2018}. Understanding the growth of seed black holes at early stages will therefore require to further assess the operation of this feedback mechanism in their specific environment. 
}

{A particularly interesting place to study the origin of intermediate-mass black holes may include low-metallicity dwarf galaxies in the metallicity range of $12+log(\mathrm{O/H})=7-9$, which may more closely resemble the conditions corresponding to the early Universe. Enhanced X-ray activity has been found in several of these \citep[e.g.][]{Prestwich2013, Reefe2023, Cann2024}, which could be due to an enhanced X-ray binary population but also an intermediate-mass black hole. These objects may include some interesting intermediate cases, as typically they do not necessarily show a fully established Nuclear Star Cluster in their center, but may include one or more Globular Cluster-like objects which potentially could be forming a dark core through the mechanism discussed above \citep[e.g.][]{Davies2011}. Within these lower-mass environments, the evolution of the cluster would not necessarily lead to the formation of an intermediate-mass black hole, but could still be contributing to the X-ray excess found in observations.  }

The results we derived here confirm important results from previous studies; particularly it has been confirmed that higher black hole formation efficiencies can be obtained when the cluster becomes relativistic, i.e. when the velocity dispersion corresponds to at least $1\%$ of the speed of light. This assumption has already been employed in the investigation by \citet{Lupi2014} in the derivation of statistical predictions from this black hole formation channel, concluding that a significant amount of seed black holes may form in this way. Our results overall confirm these conclusions and suggest that the implications of this formation channel need to be taken into account in assessments of the formation history of supermassive black holes.

\begin{acknowledgements}
      DRGS gratefully acknowledges support by the ANID BASAL projects ACE210002 and FB210003, via the Millenium Nucleus NCN19-058 (TITANs) and via Fondecyt Regular (project code 1201280). DRGS thanks for funding via the  Alexander von Humboldt - Foundation, Bonn, Germany. BR acknowledges funding through ANID (CONICYT-PFCHA/Doctorado acuerdo bilateral DAAD/62180013), DAAD (funding program number 57451854), and the International Max Planck Research School for Astronomy and Cosmic Physics at the University of Heidelberg (IMPRS-HD). MF acknowledges funding through BASAL FB210003 and ACE210002. MCV acknowledges funding through ANID (Doctorado acuerdo bilateral DAAD/62210038) and DAAD (funding program number 57600326).

\end{acknowledgements}



%
%

\end{document}